\documentstyle[prb,aps,epsfig,twocolumn]{revtex}

\renewcommand{\vec}[1]{{\bf #1}}
\newcommand{\mean}[1]{\left\langle #1 \right\rangle_{\vec{k}}}
\newcommand{\bra}[1]{\left\langle #1 \right|}
\newcommand{\ket}[1]{\left| #1 \right\rangle}
\newcommand{\operator}[1]{\hat{{\rm #1}}}
\newcommand{\unitvec}[1]{\hat{\vec{#1}}}

\newcommand{\tr}{{\rm Tr}\,}
\newcommand{\Ef}{E_{\rm F}}
\newcommand{\vf}{v_{\rm F}}
\newcommand{\Dbar}{\eta}
\newcommand{\Dpar}{D_{\|}}
\newcommand{\Dperp}{D_{\perp}}
\newcommand{\tperp}{t_{\perp}}
\newcommand{\vperp}{v_{\perp}}
\newcommand{\vpar}{v_{\|}}
\newcommand{\tauphi}{\tau_{\varphi}}
\newcommand{\tauel}{\tau_{\rm el}}
\newcommand{\lcoh}{L_{\varphi}}
\newcommand{\lel}{\l_{\rm el}}
\newcommand{\lpar}{l_{\|}}
\newcommand{\lperp}{l_{\perp}}
\newcommand{\wpar}{w_{\|}}
\newcommand{\wperp}{w_{\perp}}
\newcommand{\wparperp}{w_{\| / \perp}}
\newcommand{\Omegapar}{\Omega_{\|}}
\newcommand{\Omegaperp}{\Omega_{\perp}}
\newcommand{\sigmaBparCperp}{\sigma_{\vec{B} \perp \unitvec{z}}^{\perp}}
\newcommand{\sigmaBparCpar}{\sigma_{\vec{B} \perp \unitvec{z}}^{\|}}
\newcommand{\sigmaBperpCperp}{\sigma_{\vec{B} \| \unitvec{z}}^{\perp}}
\newcommand{\sigmaBperpCpar}{\sigma_{\vec{B} \| \unitvec{z}}^{\|}}
\newcommand{\sigmaZeroCperp}{\sigma_{0}^{\perp}}
\newcommand{\sigmaZeroCpar}{\sigma_{0}^{\|}}
\newcommand{\sigmaCperp}{\sigma^{\perp}}
\newcommand{\sigmaCpar}{\sigma^{\|}}
\newcommand{\alphaCperp}{\alpha^{\perp}}
\newcommand{\alphaCpar}{\alpha^{\|}}
\newcommand{\betaBpar}{\beta_{\vec{B} \perp \unitvec{z}}}
\newcommand{\betaBperpCperp}{\beta_{\vec{B} \| \unitvec{z}}^{\perp}}
\newcommand{\betaBperpCpar}{\beta_{\vec{B} \| \unitvec{z}}^{\|}}
\newcommand{\mal}{\raisebox{0.3ex}{\tiny $\,\times\,$}}

\begin{document}

\title{%
  Dimensional Crossover of Weak Localization in a
  Magnetic Field}

\author{C. Mauz, A. Rosch, P. W{\"o}lfle}

\address{Institut f{\"u}r Theorie der Kondensierten Materie,
  Universit{\"a}t Karlsruhe, D--76128 Karlsruhe, Germany}

\date{\today}

\twocolumn[\hsize\textwidth\columnwidth\hsize\csname 
@twocolumnfalse\endcsname %
\maketitle

\begin{abstract}
  We study the dimensional crossover of weak localization in strongly
  anisotropic systems. This crossover from three-dimensional behavior
  to an effective lower dimensional system is triggered by increasing
  temperature if the phase coherence length gets shorter than the
  lattice spacing $a$. A similar effect occurs in a magnetic field if
  the magnetic length $L_m$ becomes shorter than
  $a(\Dpar/\Dperp)^\gamma$, where $\Dpar/\Dperp$ is the ratio of the
  diffusion coefficients parallel and perpendicular to the planes or
  chains.  $\gamma$ depends on the direction of the magnetic field,
  e.g.  $\gamma=1/4$ or $1/2$ for a magnetic field parallel or
  perpendicular to the planes in a quasi two-dimensional system.  We
  show that even in the limit of large magnetic field, weak
  localization is {\em not} fully suppressed in a lattice system.
  Experimental implications are discussed in detail.
\end{abstract}


\vskip1pc]

\section{Introduction}

Strongly anisotropic systems show a number of interesting and unusual
physical properties. Many of these are due to the interplay of their
lower dimensional substructures and three-dimensional coherence.
Prominent examples are high $T_c$-superconductors in quasi two- and
Peierls' transitions in quasi one-dimensional systems. Also, weak
localization (WL) -- the subject of this paper -- qualitatively and
quantitatively depends on the effective dimension of the system.

It is important to distinguish two classes of anisotropic systems.
The first class shares more or less the properties of typical
three-dimensional systems, e.g. the Fermi surface is strongly
anisotropic but closed, and the transport properties in different
directions may vary by a large amount but still show a typical
three-dimensional behavior.

More interesting is the second class where the distance of the planes
or chains $a$ becomes an important new length scale.  As far as
electronic properties are concerned, we expect this to happen only for
systems with an open Fermi surface.  The distance $a$ should be
compared with other characteristic length scales of the system, e.g.
the phase-coherence length $\lcoh$. In the regime of low
temperatures $\lcoh$ is large, $\lcoh\gg a$, and even quasi one- or
two-dimensional systems show three-dimensional behavior.

With increasing temperature or as a function of some external
parameter like pressure or magnetic field, a dimensional crossover to
one- or two-dimensional behavior can be observed, when $\lcoh$ or an
other relevant length scale gets to be smaller than the distance $a$
of the chains or planes.

An example for this phenomenon are certain disordered quasi one- or
two-dimensional materials, which stay metallic at lowest temperatures,
despite the fact that all one or two-dimensional disordered systems
are insulators.

In this paper we will study the dimensional crossover of WL in
strongly anisotropic $3D$ systems as a function of temperature (or
phase-coherence time) and magnetic field \cite{Lee:85}. We will argue
that the magnetoresistance is an especially well suited tool to
analyze the dimensional crossover. We will first concentrate on quasi
two-dimensional systems of weakly coupled planes. The transverse and
parallel conductivity is investigated in a magnetic field either
parallel or perpendicular to the planes.  We will also mention related
results for quasi one-dimensional structures. At the end we will
discuss experimental implications and the question of universality.

This work was motivated by recent experimental results on the quasi
one-dimensional organic conductors $H_2(pc)I$ and $Ni(pc)I$, where
``$pc$'' stands for ``phthalocyanine'', which appear to show a
dimensional crossover in their WL correction to the
conductivity \cite{Lee:96}.

\section{Weak Localization and Dimensional Crossover}

WL \cite{Abrahams:79,Gorkov:79,Altshuler:85} is a quantum-interference
effect of \mbox{(particle-)} waves in a random medium. Consider the
probability for a particle to return to the position where it has
started\cite{Bergmann:83}.  For a quantum particle this
probability is given by the modulus squared of the sum of transition
amplitudes over all paths.  Due to the random nature of the phases
typically all interference effects vanish, resulting in a probability
being that of a classical diffusion process.  The cancelation of
interference terms does not hold for time reversed paths, where the
particle moves the same path clockwise and anti-clockwise, collecting
the same phase. Thus this interference is constructive, therefore
increasing the probability to return to the origin. As a consequence
the diffusion constant and the conductivity decrease. In one or two
dimensions, this so called ``coherent back-scattering'' finally leads
to localization \cite{Lee:85,Vollhardt:92}. As a magnetic field breaks
the time-reversal invariance, leading to different Aharonov-Bohm
phases for time-reversed paths, it destroys WL for all paths which
include of the order of one or more flux-quanta \cite{Lee:85}.

Technically this process can be described by the contribution of
maximally crossed diagrams to the conductivity
\cite{Lee:85,Abrahams:79,Gorkov:79}. In the hydrodynamic limit
($\vec{q} \to 0, \omega \to 0$) these contributions sum up to a
typical pole-structure, called cooperon, which (by time reversal) is
directly related to the diffusion pole \cite{Vollhardt:92}.  The
cooperon $C(\vec{r},t)$ is the solution of a diffusion equation
\cite{Altshuler:85}
\begin{equation}
  \label{cooperon1}
  \left[ \partial_t + \sum_\alpha D_{\alpha\alpha} \operator{Q}_\alpha^2 
    + \frac{1}{\tauphi} \right] C(\vec{r},t) 
  = \frac{1}{\tauel} \delta(t) \delta(\vec{r}),
\end{equation}
where we choose our coordinate system, so that the tensor of diffusion
coefficients $D_{\alpha \beta}$ is diagonal.
\begin{equation}
  \operator{\vec{Q}}=-i \hbar \vec{\nabla}-(2e/c) \operator{\vec{A}}
  \label{momentum-operator}             
\end{equation}
is the canonical momentum operator.  We include the vector potential
$\vec{A}$ throughout the paper in the definition of
$\operator{\vec{Q}}$ via minimal coupling in a semi-classical
approximation.  Note the factor of two in front of the charge, as the
cooperon describes a two-particle process.  The fact, that the
particle looses its phase coherence due to electron-electron or
electron-phonon interaction is described by the phase coherence time
$\tauphi$, which acts as an infrared-cutoff for (\ref{cooperon1}).
Only processes on time-scales shorter than $\tauphi$ contribute to WL.
Generally $\tauphi$ and the elastic scattering time $\tauel$ will be
$\vec{k}$-dependent especially in the strongly anisotropic systems we
want to describe.  However, we will neglect this for simplicity, as it
should not affect our qualitative conclusions.

The correction $\Delta \sigma$ due to the cooperon to the Boltzmann
conductivity $\sigma_{0,\alpha\beta}$ is given by
\begin{eqnarray}
  \Delta\sigma_{\alpha\beta}&=&
  e^{2} \int \frac{d\vec{k}}{(2\pi)^{3}}
  \int \frac{d\vec{Q}}{(2\pi)^{3}}\nonumber\\
  && \mal v_{\alpha}(\vec{k}) |G_{\vec{k}}|^{2}
  C(\vec{Q},\omega) 
  |G_{\vec{Q}-\vec{k}}|^{2} v_{\beta}(\vec{Q}-\vec{k});
\end{eqnarray}
$\vec{v}(\vec{k})=\partial_{\vec{k}} \epsilon_{\vec{k}}$ is the
velocity. The Green's functions $G_{\vec{k}}$ are sharply peaked in a
region $\cal F$ of width $1/\tauel$ around the Fermi surface
($1/\tauel < \Ef$).  Therefore, the $\vec{k}$-summation can be
replaced by an average over the region $\cal F$, which we denote by
$\mean{\dots}$ and
\begin{equation}
  \label{eq:WL:correction}
  \Delta\sigma_{\alpha\beta}=\frac{e^{2}\tauel^{2}}{\pi} 
  \int \frac{d\vec{Q}}{(2\pi)^{3}}
  \mean{v_{\alpha}(\vec{k}) v_{\beta}(\vec{Q-k})} C(\vec{Q},\omega).
\end{equation}
To discuss the influence of a magnetic field it is essential to
incorporate the vector potential in a gauge-invariant way. We
therefore rewrite (\ref{eq:WL:correction}) in the manifestly
gauge-invariant form
\begin{equation}
  \label{eq:correction:gaugeinvariant}
  \Delta\sigma_{\alpha\beta}=\frac{e^{2}\tauel^{2}}{\pi V} 
  \tr \!\left[ 
   \mean{ 
     \operator{v}_{\beta}(\vec{Q-k}) 
     \operator{v}_{\alpha}(\vec{k})} 
   \operator{C}(\vec{Q},\omega)
  \right]_{\vec{Q}},
\end{equation}
where the averaged velocities and the cooperon are understood as
operators (we denote operators with a hat) in the center-of-mass space
spanned by $\hat{\vec{Q}}$ or its conjugate position operator
$\hat{\vec{r}}$. $V$ is the volume of the system.

When the average of the velocities is $\vec{Q}$ independent, as it is
e.g. in an isotropic $3D$ system, using
$\sigma_{0,\alpha\beta}(\omega=0,B=0)= e^2 \mean{v_{\alpha}
  v_{\beta}}=N_{0}e^{2}D_{\alpha\beta}$ for the static Boltzmann
conductivity, formula (\ref{eq:WL:correction}) reduces to the well
known form
\begin{equation}
  \label{delsigma}
  \frac{\Delta \sigma}{\sigma_0}
  =-\frac{\tauel}{\pi N_0} C(\vec{r}=0,\omega=0),
\end{equation}
$N_0$ is the density of states at the Fermi-surface.

For the discussion of a dimensional crossover, it is
essential to incorporate the spatial structure and the anisotropy of
the system in the diffusion equation for the cooperon
(\ref{cooperon1}).  The range of validity of the diffusion equation
(\ref{cooperon1}) for small distances is on the one hand restricted by
the elastic scattering length $\l_\alpha=\sqrt{D_{\alpha \alpha}
  \tauel}$, on the other by the structure of the material, i.e. by the
spacing $a$ of the planes in the anisotropic quasi $2D$ system.
Therefore one has to distinguish the two scenarios mentioned in the
introduction: in the first scenario, the elastic scattering length
perpendicular to the planes $\lperp$ is larger than the distance $a$
of the planes.  Then the ultraviolet cutoff in the
$q_{\perp}$-integration is given by $\lperp^{-1}$ and the anisotropy
of the diffusion-constant in (\ref{cooperon1}) can be scaled out as
previously shown \cite{Woelfle:84}. Such a material will always show
$3D$ behavior, as far as WL is concerned.  We expect this first
scenario, e.g.  in all materials with a closed Fermi-surface, which
can continuously be deformed to a sphere. The electronic properties of
such systems are qualitatively not directly influenced by their lower
dimensional substructures.

The second scenario is the more interesting one. Here $\lperp$ is
formally smaller than $a$ and the cutoff is now given by the
substructure of the system, i.e.  the distance of the planes.  As we
will see, these systems show a dimensional crossover and are
affected by the underlying lattice structure. We expect this scenario
to hold for anisotropic systems with an open Fermi-surface and quite
strong short-range disorder.

We measure the anisotropy with the dimensionless constant
\begin{equation}
  \Dbar=\frac{\Dperp \tauel}{a^2}=
  \left(\frac{\lperp}{a}\right)^2.
\end{equation}
For $\Dbar \ll 1$ we are in the second regime. In terms of microscopic
quantities the diffusion constants are given by $\Dperp = \tperp^2 a^2
\tauel/2$ or $\Dbar=(\tperp\tauel)^2/2 $ for an effective hopping
rate $\tperp$ perpendicular to the planes. The diffusion constant for
the motion in the planes is $\Dpar=\vf l/2$ where $l=\vf \tauel$ is
the (elastic) mean free path and $\vf$ the Fermi velocity. A
dimensional crossover should therefore be observable if the broadening
due to elastic scattering is larger than the bandwidth $\tperp$
perpendicular to the planes.

As it turns out, in the case of a finite magnetic field it is
important to include the cutoffs of the cooperon-equation
(\ref{cooperon1}) in a gauge-invariant way.  Therefore we will analyze
the following cooperon-equation, here given for a quasi-$2D$ material
with the symmetry axis in the $z$-direction and frequency $\omega=0$:
\begin{equation}
  \left[ \Dpar \left(
      \operator{Q}_{x}^2+ \operator{Q}_{y}^2\right) 
    +\frac{4\Dperp}{a^2} \sin^{2}\frac{a\operator{Q}_z}{2}  
    +\frac{1}{\tauphi}\right] C(\vec{r})
  =\frac{1}{\tauel} \delta(\vec{r}).
  \label{cooperon2}
\end{equation}
Eq. (\ref{cooperon2}) is obtained from the result of summing the
maximally crossed diagrams in zero magnetic field for a tight-binding
model description of weakly coupled planes
\cite{Prigodin:84,Nakhmedov:86,Dupuis:92a} by
substituting the canonical momentum operator (\ref{momentum-operator})
for the Cooperon momentum $\operator{\vec{Q}}$ in a quasi-classical
approximation.

The ''bandstructure'' $2 \sin^{2} (a Q_z/2)=1-\cos a Q_z $ serves as a
convenient physical way to introduce the cutoff $a$.  Note that the
details of the cutoff structure, i.e. the special form of the
electronic bandstructure, will not affect any of our results
qualitatively.  In the $z$-direction the diffusion described by Eq.
(\ref{cooperon2}) takes place on a lattice with spacing $a$.  For
convenience we choose a gauge with $A_z=0$ in (\ref{cooperon2}),
otherwise the usual Peierls' phase factors for a magnetic field on a
lattice have to be introduced.

Dupuis and Montambaux \cite{Dupuis:92a} investigated the validity of
the diffusion approximation for large magnetic fields analyzing the
exact integral kernel for the cooperon.  They concluded that
(\ref{cooperon2}) is valid for not too strong magnetic field
$\omega_{c}\tauel\ll 1$.  The cyclotron-frequency $\omega_{c}$ is
given by $\omega_{c}=e B /mc$ for a magnetic field parallel to the
planes (closed orbits) and by $\omega_{c}\approx eB\vf a/c$ in the
case of a magnetic field perpendicular to the planes (open orbits). A
diffusion equation for the cooperon similar to (\ref{cooperon2}) was
analyzed by Nakhmedov {\it et al.}\cite{Nakhmedov:86} to study the
dimensional crossover from $d=1$ to $d=2$ two-dimensional system.
Dorin \cite{Dorin:93} investigated quasi two-dimensional systems but
considered only a magnetic field perpendicular to the planes. Dupuis
and Montambaux \cite{Dupuis:92a} studied the conductivity in a
strongly anisotropic two-dimensional system, but also for anisotropic
three-dimensional materials. They focused their attention to the
in-plane conductivity (see also
Ref.~\onlinecite{Cassam-Chenai:94}).  We will argue
below that for an experimental analysis of the effects of weak
localization in a magnetic field, the comparison of in-plane and
out-of-plane conductivity for both a magnetic field parallel and
perpendicular to the planes is necessary. We will therefore discuss
all four situations in detail for both small and large magnetic fields
in the whole regime where (\ref{cooperon2}) is valid.

\section{Quasi Two-Dimensional Systems}
\subsection{Perpendicular Field}
We first consider the effect of a magnetic field perpendicular to the
planes of a quasi two-dimensional system. To model the system we
assume an open Fermi-surface with a band-structure of the type
$\epsilon_{\vec{k}}=(k_x^2+k_y^2)/(2 m)-2 \tperp \cos(a k_z)$ with
$\tperp < 1/\tauel< \Ef$. The details of $\epsilon_{\vec{k}}$ do not
affect any of our qualitative results as long as the topology of the
Fermi-surface remains unchanged.

For a magnetic field perpendicular to the planes the magnetic field
only affects the motion in the planes. This situation was recently
analyzed by Dorin \cite{Dorin:93}. In this section we will more or
less rederive his results and establish notations and techniques for
the more involved discussion of a magnetic field parallel to the
planes.

\subsubsection{Current in the planes}
For the case of a current directed parallel to the planes the
velocity average over the region $\cal F$ is given by
\begin{equation}
  \mean{\vpar(\vec{k}) \vpar(\vec{Q-k})}=-\frac{1}{2}\vf^{2}.
  \label{mean:parallel}
\end{equation}
To calculate the WL correction to the in-plane conductivity $\Delta
\sigmaBperpCpar$, the trace in (\ref{eq:correction:gaugeinvariant}) is
expanded in normalized eigenfunctions $\Phi_\lambda$ of the
 operator on the left-hand side of (\ref{cooperon2})
for $\omega=0$ defined by
\begin{equation}\label{eigenfunctions}
  \left[ \Dpar \left( {Q}_{x}^2+ {Q}_{y}^2\right)
    +\frac{4\Dperp}{a^2} \sin^{2}\frac{a{Q}_z}{2}
    +\frac{1}{\tauphi}\right] \Phi_{\lambda}
  = E_{\lambda}\Phi_{\lambda}.
\end{equation}
As the velocity average is just a constant, $\Delta \sigmaBperpCpar$
can be expressed in terms of the eigenvalues $E_\lambda$
\begin{equation} 
  \label{efentw}
  \frac{\sigmaBperpCpar}{\sigma_0}=-\frac{1}{\pi N_0 V} \sum_\lambda 
  \frac{1}{E_\lambda}
\end{equation}
with $N_0\approx m/(2 \pi a)$.  The ultra-violet cutoff can be imposed
in a gauge invariant way by requiring $E_{\lambda}\lesssim 1/\tauel$.

For the calculation we choose the gauge $\vec{A}_{\perp}=(0,B x,0)$.
The strength of the magnetic field is measured by the magnetic length
$L_m=\sqrt{\hbar c/eB}$.  The eigenfunctions $\Phi_\lambda$ can be
decomposed in plane waves in the $y$- and $z$-direction carrying
momenta $q_{y}$ and $q_{z}$ and a remaining part, which describes a
shifted one-dimensional harmonic oscillator (HO) for a particle of
``mass'' $(2\Dpar)^{-1}$ and ``cyclotron frequency''
$\Omegaperp=4\Dpar eB/c=4 \Dpar/L_m^2 $ with eigenfunctions
$\Psi^{\rm HO}_n(x+L_{m}^{2}q_{y}/2)$ and eigenvalues
$E'_{n}=(n+1/2)\Omegaperp$, $n=0,1,2,\dots$. Note that $\Omegaperp$ is
by a factor of $4 \Dpar m \approx 4 \Ef \tauel$ larger than the
cyclotron frequency of bare electrons.  Therefore a weak magnetic
field is sufficient to have a strong effect on WL.

The DC-conductivity is found from (\ref{efentw}) as
\begin{equation}\label{summ0}
  \frac{\Delta \sigmaBperpCpar}{\sigmaZeroCpar}=
  \frac{-1}{L_m^2 \pi^2 N_{0}}  \sum_{n, q_z}
  \left({E_n'+ \frac{4\Dperp}{a^2}
    \sin^{2} \frac{a q_z}{2} + \frac{1}{\tauphi}}\right)^{-1}.
\end{equation}
The factor $2/(2 \pi L_m^2) $ takes care of the degeneracy of the
``Landau levels'' of the cooperon with charge $2 e$.  For $\Dbar<1$,
$q_z$ is integrated over the whole Brillouin zone $|q_{z}| < \pi/a$
leading to
\begin{equation}\label{summ1}
  \frac{\Delta \sigmaBperpCpar}{\sigmaZeroCpar}= 
  -\lambda \hspace*{-1.5em}\sum_{\epsilon_{n}=(n+\frac{1}{2})\wperp}
  \frac{\wperp \, g_c(\epsilon_{n})}%
  {\sqrt{\epsilon_{n}+x}\sqrt{\epsilon_{n}+x+4\Dbar}}.
\end{equation}
Here we have introduced the dimensionless quantities
\begin{equation}\label{dimensionlessPerp}
  x=\frac{\tauel}{\tauphi}, \quad 
  \lambda=(2\pi\Ef\tau)^{-1},  \quad 
  \wperp=\Omegaperp\tauel=\frac{4\lpar^{2}}{L_m^{2}}
\end{equation}
to measure the rate of phase-destruction, the strength of disorder and
the cyclotron frequency. Sometimes it is useful to interpret $\wperp$
as the inverse area occupied by a flux quantum in units of a typical
area, where elastic processes take place. $g_c(x)$ is a
cutoff-function of scale unity --- for our plots we use
$g_c(x)=\exp(-\pi x^{2}/4)$ whereas our analytical formulas are given
for the step function $g_c(x)=\Theta(1-x)$.

\begin{figure}
  \begin{center}
    \epsfig{width=\linewidth,file=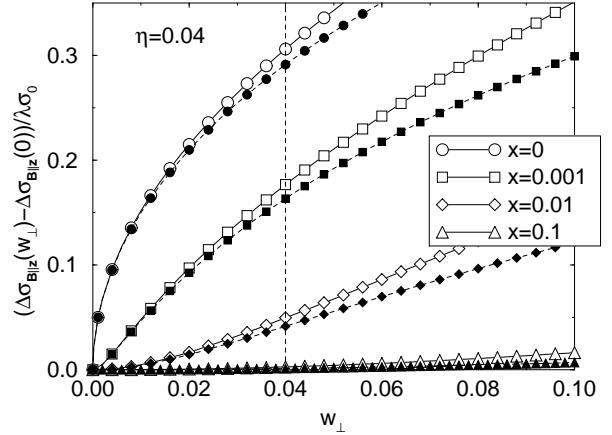}
    \caption{The magneto-conductivity for a field perpendicular
      to the planes as a function of magnetic field $\wperp=4
      \lpar^2/L_{m}^{2}$ for various temperatures, i.e. various phase
      coherence times $x=\tauel/\tauphi$.  The open and
      filled symbols denote the conductivity parallel and
      perpendicular to the planes, respectively. The crossover scale
      $\wperp=\Dbar=0.04$
      is shown as a dashed line. The lines are guides to the eye.}
    \label{fig:1}
  \end{center}
\end{figure}

For weak magnetic field $\wperp \to 0$, the sum (\ref{summ1}) can be
approximated by an integral, and the effects of discretization can be
calculated using the Euler-Maclaurin summation formula. For the
different regimes we find the following analytical approximations
\begin{eqnarray} 
  \label{cases}
  \frac{\Delta \sigmaBperpCpar}{\sigmaZeroCpar}&\approx 
  &-\lambda\left\{
    \begin{array}{l@{\quad}l}
      \alphaCpar-\betaBperpCpar\wperp^2,  & \wperp \ll x, \Dbar \\
      \displaystyle\alphaCpar-c \sqrt{{\wperp}/{\Dbar}}, 
      & x \ll \wperp \ll \Dbar \\
      \displaystyle\gamma+\log ({4}/{\wperp}),  
      & x, \Dbar \ll \wperp \ll 1  \\
      2 g_c(\wperp/2), & \wperp \gg 1\\
      \propto 1/x, & x \gg 1
    \end{array}
  \right.
\end{eqnarray}
with Euler's Gamma constant $\gamma$, a constant $c =
(\zeta(1/2)/2)(1-\sqrt{2}) \approx 0.3024$ and
\begin{eqnarray}
  \alphaCpar(\Dbar,x)&=&\int_0^\infty  
  \frac{g_c(\epsilon)}{\sqrt{\epsilon+ x }\sqrt{\epsilon+x+4 \Dbar}}
  \,d\epsilon \nonumber\\
  &=& \kappa_{1}(x+1)-\kappa_{1}(x)
  \label{eq:alphaCpar}\\
  \betaBperpCpar(\Dbar,x)&=&-(2\Dbar+z)\lambda_{1}(z)\Big|_{z=x}^{x+1}
  \approx (2\Dbar+x)\lambda_{1}(x).
  \label{eq:betaBperpCpar} 
\end{eqnarray}
We have introduced for later convenience 
the functions $\kappa_{1},\lambda_{1}$
\begin{eqnarray}
  \kappa_{1}(z)&=&2 \log \left(\sqrt{z}+\sqrt{z+4\Dbar}\right)
  \label{eq:kappaOne}\\
  \lambda_{1}(z)&=&\frac{1}{24 z^{3/2}(z+4\Dbar)^{3/2}}
  \label{eq:lambdaOne}.
\end{eqnarray}
A graphical representation based on a numerical evaluation of
(\ref{summ1}) is given in Fig. \ref{fig:1}.
In the limit $\Dbar \to 0$ one recovers the well known $2D$ results
\cite{Lee:85}.  The precise form of $\alphaCpar(\Dbar,x)$ and the
constant contribution in the regime $x,\Dbar \ll \wperp \ll 1$ depend
slightly on the cutoff structure, i.e. on $g_c$.

Using (\ref{eq:alphaCpar}), we can now discuss the dimensional
crossover as a function of temperature in the absence of a magnetic
field.  To a good approximation the only temperature dependent
quantity in our model is the phase-coherence time contained in the
parameter $x=\tauel/\tauphi$. Usually its temperature dependence can
be approximated by a power law $x \propto T^{\gamma}$, where $\gamma$
depends on the dominant scattering process \cite{Lee:85,Schmid:73}.

For $x<\Dbar$, i.e. if the phase-coherence length $\sqrt{\Dperp
  \tauphi}$ is larger than the distance of the planes $a$, the usual
three-dimensional behavior is observed. For not too strong disorder
the system is metallic as the correction due to WL is finite for $T\to
0$, i.e. $x \to 0$.  The leading temperature dependence is given by
the contribution proportional to $\sqrt{x /\Dbar}$ -- again typical
for WL in a $3D$ system. However, at higher temperatures the coherence
between the planes is destroyed and the behavior of the system is
dominated by the two-dimensional planes.  As a consequence, we obtain
the logarithmic correction $\Delta \sigma/\sigma_{0} \approx -\lambda
\log x$ typical for WL in two dimensions.

The magneto-conductivity at low temperature rises quadratically with
the magnetic field for $\wperp \ll x$, i.e. as long as a particle can
not enclose a flux quantum in a coherent diffusion process. For a
higher field, the magneto-conductivity (cf.  Fig.~\ref{fig:1})
increases with the square-root of the magnetic field in the
three-dimensional regime with $x \ll \wperp \ll \Dbar$. Note that
corrections to the square-root dependence are of the order of
$x/\sqrt{\wperp \Dbar}$, therefore this regime can only be observed at
quite low temperatures.  By increasing the magnetic field even more,
the WL corrections crosses over to a logarithmic dependence on field.
Finally, at a very high field when $\wperp \gg 1$, WL is destroyed
completely (not shown in Fig.~\ref{fig:1}).

\subsubsection{Current perpendicular to the planes}
For the case of the current perpendicular to the planes, the velocity
average is now $\vec{Q}$ dependent
\begin{eqnarray}
  \mean{\vperp(\vec{k}) \vperp(\vec{Q-k})}
  &\approx&-\frac{1}{2}\tperp^{2}a^{2} \cos(a Q_{\perp})\nonumber\\
  &=&-\frac{\Dperp}{\tauel} \cos(a Q_{\perp}).
  \label{mean:perpendicular}
\end{eqnarray}
While (\ref{mean:perpendicular}) depends on the details of the
band-structure, our results are nearly independent of these details.
Our conclusions rely on the generic feature that
$\mean{\vperp(\vec{k}) \vperp(\vec{Q-k})} \to const.$ for $Q_\perp\to
0$ and on $\int d Q_\perp \mean{\vperp(\vec{k}) \vperp(\vec{Q-k})}=0$.
The last fact is a consequence of the periodicity of
$\epsilon_{\vec{k}}$.

In the chosen gauge ($A_{\perp}=(0,0,By)$), $Q_{\perp}=q_z$ is a good
quantum number.  Therefore it is easy to add the factor $\cos a q_z$
in (\ref{summ0}) and the WL correction perpendicular to the planes has
the form\cite{Dorin:93}
\begin{eqnarray}
  \frac{\Delta \sigmaBperpCperp}{\sigmaZeroCperp}&=&
  -\frac{\lambda}{4\Dbar} \sum_{\epsilon_{n}=(n+\frac{1}{2})\wperp}
  \frac{\wperp \, g_c(\epsilon_{n})}%
  {\sqrt{\epsilon_{n}+x}\sqrt{\epsilon_{n}+x+4\Dbar}}\nonumber\\
  && \qquad \mal \left(
    \sqrt{\epsilon_{n}+x+4\Dbar}-\sqrt{\epsilon_{n}+x} \right)^{2}.
\end{eqnarray}
Again, analytical approximations can be found in the different regimes:
\begin{eqnarray}
    \frac{\Delta \sigmaBperpCperp}{\sigmaZeroCperp}
    &\approx& -\lambda
    \left\{
      \begin{array}{l@{\quad}l}
        \alphaCperp-\betaBperpCperp \wperp^{2}, & \wperp \ll x,\Dbar\\
        \alphaCperp-c\sqrt{{\wperp}/{\Dbar}}, 
        & x \ll \wperp \ll \Dbar\\
        {(\Dbar\pi^{2})}/{(2 \wperp)}, &  x, \Dbar \ll \wperp \ll 1\\
        \frac{4\Dbar}{\wperp}g_{c}\left({\wperp}/{2}\right), 
        & \wperp \gg 1\\
        \propto 1/x^{2}, & x\gg 1
      \end{array}
    \right.
  \end{eqnarray}
with the numerical constant $c\approx 0.30$ as before,
\begin{eqnarray}
  \alphaCperp(\Dbar,x)&=&\frac{1}{2\Dbar}
  \left(\sqrt{z(z+4\Dbar)}-z\right)\Big|_{z=x}^{1+x} \label{eq:alphaCperp}\\
  \betaBperpCperp(\Dbar,x)&=&-2\Dbar \lambda_{1}(z)\Big|_{z=x}^{x+1}
  \approx 2 \Dbar \lambda_{1}(x) .
\end{eqnarray}
$\lambda_{1}$ is defined by Eq. (\ref{eq:lambdaOne}).

For small fields and low temperatures $\wperp, x \ll \Dbar$, the main
contribution is due to the Cooper-pole with $\vec{Q}\approx 0$ and
therefore $\cos a q_z\approx 1$.  As a consequence we get practically
the same answers for both directions of the current as can be seen in
Fig. \ref{fig:1}. However, at high fields $\wperp \gg \Dbar$ (not
shown in Fig.  \ref{fig:1}, see Fig. \ref{fig:2}) or temperatures
$x \gg \Dbar$ the WL correction $\Delta \sigmaBperpCperp$
vanishes faster because the coherence among different planes is
destroyed.  A crossover from three- to zero-dimensional behavior is
observed \cite{Dorin:93}.

\subsection{Parallel Field}
\subsubsection{Current in the planes}
More interesting is a magnetic field parallel to the planes.
We will first discuss the case of a current in the planes where the
average of the velocities is given by (\ref{mean:parallel}).  We
choose the gauge $\vec{A}_{\|}=(0,-B z,0)$ for a magnetic field in
$x$-direction.  Again, the eigenfunctions defined in
(\ref{eigenfunctions}) can be decomposed in plane waves in $x$ and $y$
direction with momenta $q_x$ and $q_y$ and a non trivial part.
Therefore we have to analyze the remaining ``Hamiltonian'':
\begin{equation}
  \operator{H}'=
  \Dpar \left(\operator{q}_y+\frac{2 \operator{z}}{L_m^2}\right)^2 
  +\frac{2\Dperp}{a^2} \left(1- \cos a \operator{q}_z \right).
  \label{cooperon3}
\end{equation}
This is the Hamiltonian of a one-dimensional tight-binding lattice
with hopping-rate $ \Dperp/a^2$ and an external harmonic potential,
centered at $-q_y L_m^2/2$, with oscillator frequency
$\Omegapar=4\sqrt{\Dperp\Dpar}/L_{m}^{2}=\Omegaperp\sqrt{\Dperp/\Dpar}$.
The corresponding ``Schr\"odinger equation'' can be identified with
Mathieu's equation \cite{Abramowitz:65,Nakhmedov:86}.

\begin{figure}
    \begin{center}
    \epsfig{width=\linewidth,file=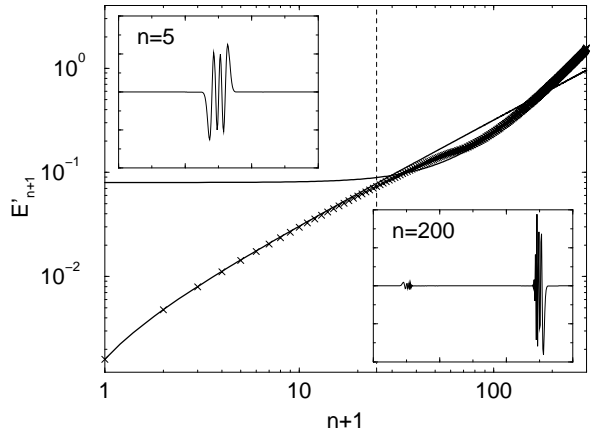}
    \caption{
      Numerically calculated spectrum (crosses) of the ``Hamiltonian''
      (\ref{cooperon3}) with $q_y=0$ for a lattice with $300$ sites in
      a log-log plot.  For small quantum numbers, the dispersion is
      linear, whereas for high quantum numbers, it crosses over to a
      quadratic dependence.  The solid lines represents our
      approximations $\epsilon_{n}=(n+1/2)\wpar$ or
      $\epsilon_{n}=2\Dbar+(n/2)^{2} \wpar^{2}/(4 \Dbar)$,
      respectively. For the plot we have chosen $2\Dbar=0.04$ and
      $\wpar=0.0032$.  In the insets, we show typical eigenfunctions
      in both regimes.  The crossover scale $n_{c}=2\Dbar/\wpar$ is
      shown as a dashed line.}
    \label{fig:3}
  \end{center}
\end{figure}

For low quantum numbers and low energies, only the quadratic part of
the kinetic energy of $ H'$ is probed, therefore the wave-functions
have approximately HO from as above. Their energy is given by $E_n'
\approx \Omegapar(n+1/2)$. This contribution will dominate for a weak
magnetic field.  We can estimate the range of validity from the
condition $q_{z}\lesssim 1/a$ using a characteristic momentum
$q_{z,typ}\approx \sqrt{2mn\Omegapar}$ and get $n\lesssim n_c=c'
\Dbar/\wpar$, where $c'$ is a constant of order one. A comparison with
numerical results suggests the value $c' \approx 2$.  We have
introduced as above the dimensionless constant $\wpar=\Omegapar
\tauel=4\lperp \lpar/L_m^2$.  In Fig. \ref{fig:3} we compare the
numerically calculated spectrum of $H'$ with this analytical
approximation.  The contribution of these lowest eigenvalues is
approximately given by
\begin{eqnarray}
  &\displaystyle 
  \left(\frac{\Delta\sigmaBparCpar}{\sigmaZeroCpar}\right)_{\rm HO}
  =-\lambda K_1& \\
  &\displaystyle K_{1} = \frac{\wpar}{\pi \sqrt{\Dbar}} 
  \sum_{n=0}^{\infty} 
  \frac{g_c\left[\left(n+\frac{1}{2}\right)\frac{\wpar}{2\Dbar}\right] 
    \arctan \left(\epsilon_{n}+x\right)^{-1/2}}{\sqrt{\epsilon_{n}+ x }}.&
  \label{K1}
\end{eqnarray}
Note the factor $2\Dbar$ in the cutoff-function taking into account the
condition $n<n_c$. The energies are given by $\epsilon_n=(n+1/2)
\wpar$. Actually it turns out that for $\wpar \approx \Dbar$, i.e. in
the crossover region, the main contribution to the
magneto-conductivity is due to the lowest eigenstate $\epsilon_0$.  In
the same region the approximation $\epsilon_0 \approx \wpar/2$ breaks
down.  Therefore we use for our numerical evaluations of (\ref{K1})
a more accurate expression for the ground-state energy
\begin{equation}
  \epsilon_0=2 \Dbar 
  \frac{8u^4+7u^2+16u}{8u^4+15u^3+16u+32}, 
  \qquad u=\frac{\wpar}{2 \Dbar}. 
  \label{eq:epsilon:Zero}
\end{equation}
This interpolating formula describes the ground state energy of
(\ref{cooperon3}) exactly in next-to-leading order\cite{Abramowitz:65}
in both $\wpar/\Dbar$ and $\Dbar/\wpar$. Especially, for $\wpar \to 0$
it reduces to $\epsilon_0\approx \wpar/2$ and for large magnetic field
to $\epsilon_0\approx 2 \Dbar$.

For high quantum numbers $n \gtrsim n_{c}$, the potential energy
dominates and the hopping rate is small compared to the
level-splitting due to the external potential.  In this regime the
eigenfunctions are localized in one plane and we can approximate
\begin{equation} 
  \operator{H}'\approx \Dpar
  \left(\operator{q}_y+\frac{2\operator{z}}{L_m^2}\right)^2 
  + \frac{2\Dperp}{a^2}.
  \label{locLoes}
\end{equation}
$\operator{z}$ denotes the position of the planes and has the
eigenvalues $n a$, $n=0,\pm 1, \dots$. Therefore the higher
eigenvalues of $\operator{H}'$ are approximately given by
\begin{equation}
  E'_n\approx \Dpar \left(q_{y}+\frac{2}{L_{m}^{2}}na\right)^{2}
  + \frac{2\Dperp}{a^2}.
\end{equation}

That this approximation is very accurate, is shown in Fig.~\ref{fig:3}
where we compare our approximation to a numerically determined
spectrum of $\operator{H}'$.  As the crossover regime will depend in
any case on the (often unknown) details of band-structure and
scattering rates and will influence our results not qualitatively, we
will just sum up the contributions from the two regimes with the
proper cutoff, denoted by the dashed line in Fig.~\ref{fig:3}.

Thus from (\ref{efentw}) we get the following contribution from the
second regime: 
\begin{equation}
  \left(\frac{\Delta\sigmaBparCpar}{\sigmaZeroCpar}\right)_{\rm loc}=
  -\lambda (K_{2}-K_{3})
\end{equation}
with 
\begin{eqnarray}
  K_{2}&=&\int_{0}^{\infty} 
  \frac{g_{c}(q^{2})}{q^{2}+2\Dbar+x} dq^2
  \approx\log \frac{2\Dbar+x+1}{2\Dbar+x}  \label{K2} \\
  K_{3}&=&\frac{\wpar}{\pi \sqrt{\Dbar}} 
  \sum_{n=0}^{\infty}
  \frac{
    g_c \left[\left(n+\frac{1}{2}\right)\frac{\wpar}{2 \Dbar}\right] 
    \arctan \left(2\Dbar+x\right)^{-1/2}}{\sqrt{ 2\Dbar + x }}\\
  &\approx& \frac{1}{\sqrt{4\Dbar(2\Dbar + x )}}
  \mal \left\{
    \begin{array}{l@{\quad}l}
      2 \Dbar,& \wpar \ll \Dbar\\
      \displaystyle \wpar g_{c} \left(\frac{\wpar}{4 \Dbar}\right),
      & \wpar \gg \Dbar 
    \end{array}
  \right. 
  .\label{K3}
\end{eqnarray}
$K_2$ is the contribution one obtains from a summation over {\em all\ 
  } eigenvalues of (\ref{locLoes}). It is independent of the magnetic
field.  As (\ref{locLoes}) does describes only the eigenstates of
(\ref{cooperon3}) for $n>n_c=2 \Dbar/\wpar$, we have to subtract the
first $n_c$ eigenvalues summing up to $K_3$.

A direct interpretation of (\ref{K2}) can be given: As long as the
particle moves in the plane, no flux is enclosed as the magnetic field
is parallel to the motion. Therefore we find a contribution
independent of the magnetic field. For $\Dbar=0$, i.e. in the true
$2D$ situation, we recover the usual two-dimensional result as
$K_{2}=\alphaCpar(\Dbar=0,x)=\log (1+1/x)$ and $K_1=K_3=0$.  However,
for a finite diffusion rate perpendicular to the planes there is an
infrared cutoff due to the fact that the particle eventually leaves
the particular plane it has been moving in. $K_2$ is exactly the
formula describing WL in a two-dimensional system, however $1/\tauphi$
is replaced by $1/\tauphi+2/\tau_{a}$, where $\tau_{a}= \tauel/\Dbar
=a^2/\Dperp$ is the time needed to diffuse to a neighboring plane.  As
a result, for very low temperatures $K_2$ saturates at $K_2(x=0)
\approx \log 1/\Dbar$, instead of diverging as for a $2D$ system.

To discuss the magnetoresistance for a finite magnetic field parallel
to the planes we have to add the contribution from the HO- and the
localized regime
\begin{equation}
  \frac{\Delta \sigmaBparCpar}{\sigmaZeroCpar} 
  \approx -\lambda (K_1+K_2-K_{3}).
\end{equation}
An analytic evaluation yields the following expression in the various
regimes
\begin{equation}
   \frac{\Delta \sigmaBparCpar}{\sigmaZeroCpar}\approx 
 -\lambda \left\{
    \begin{array}{l@{\quad}l}
      \alphaCpar-\betaBpar\wpar^{2}, & \wpar \ll x,\Dbar \\
      \alphaCpar-c \sqrt{{\wpar}/{\Dbar}}, & x \ll \wpar \ll \Dbar \\
      \displaystyle K_2, 
      & \wpar \gg \Dbar,x\\
      \propto 1/x, & x \gg 1
    \end{array}    
  \right.\label{K1ana}
\end{equation}
with
\begin{eqnarray}
  \betaBpar&=&\lambda_{2}(x+2\Dbar)-\lambda_{2}(x)\\
  \lambda_{2}(z)&=&-\frac{1}{48\pi\sqrt{\Dbar}}
  \left[ z^{-3/2}\arctan\frac{1}{\sqrt{z}}+\frac{1}{z(1+z)} \right].
\end{eqnarray}
As in the case of a perpendicular field, the (negative)
magnetoresistance rises quadratically with magnetic field for $\wpar
\ll x,\Dbar$ and then crosses over to a square-root dependence on $B$
or $\wpar$.

In contrast to the usual three-dimensional situation or a magnetic
field perpendicular to the planes, the WL correction for magnetic
field parallel to the planes does {\em not} vanish in a large magnetic
field.  The contribution $K_2$ remains finite even at very high
fields\cite{Nakhmedov:86,Dupuis:92a} as explained above.  

At very high fields, $\omega_c \tau \gg 1$ or $\wpar \gg \sqrt{\Dbar}$
the diffusion approximation (\ref{cooperon2}) breaks down as has been
investigated by Dupuis and Montambaux \cite{Dupuis:92a}.  They find
that for very high fields $\wpar \gg 1$ the wave function of the {\em
  electrons} is localized in the planes leading to a strong
suppression of hopping to neighboring planes. As the motion in the
planes is not affected by a parallel magnetic field the effect of weak
localization increases for a stronger magnetic field in this regime.
As a consequence a {\em positive} magnetoresistance is expected for
theses extremely high fields.
\begin{figure}
  \begin{center}
    \epsfig{width=\linewidth,file=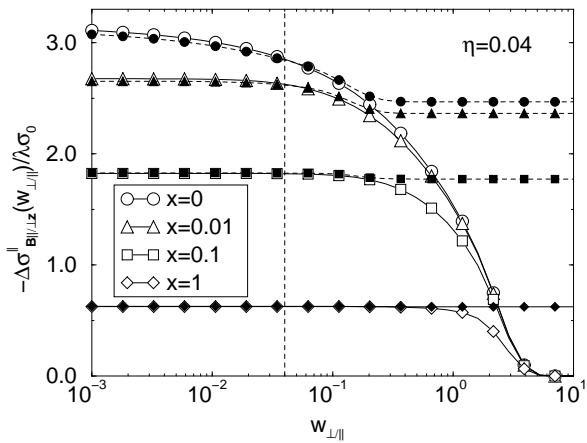}
    \caption{Comparison of the WL correction of the in-plane
      conductivity for a magnetic field parallel to a field
      perpendicular to the planes for various phase coherence times
      $x=\tauel/\tauphi$, $\Dbar=0.04$.  Note that on the $x$-axis
      different scales have been used: $\wpar=4\lperp \lpar/L_m^2$ for
      a magnetic field in the planes (filled symbols) and $\wperp=4
      \lpar^2/L_m^2= \wpar \sqrt{\Dpar/\Dperp}$ for a magnetic field
      in the perpendicular direction (empty symbols).  The crossover
      scale $w_{\perp/\|}\sim \Dbar$ is shown as a dashed line. The
      lines are guides to the eye.}
    \label{fig:2}
  \end{center}
\end{figure}
In Fig. \ref{fig:2}, the effects of a magnetic field  parallel and
perpendicular to the planes are compared. In the low temperature
regime $x\ll\Dbar$, the dimensional crossover is induced by the
magnetic field for $\wperp \sim \Dbar$ (field perpendicular to
the planes) or $\wpar\sim \Dbar$ (field parallel to the planes).
For higher temperatures $x \gtrsim \Dbar$, such an crossover can not be
observed because it is already preempted by the phase destroying
scattering. At very high magnetic field $\wpar \gg \Dbar$, a finite
contribution of WL remains when the field is parallel to the planes.

\subsubsection{Current perpendicular to the planes}
To calculate the out-of-plane WL-correction to the conductivity, we
have to analyze (\ref{eq:correction:gaugeinvariant}) using the
velocity average (\ref{mean:perpendicular}). Expanding the trace in
(\ref{eq:correction:gaugeinvariant}) in eigenfunctions
$\ket{\Phi_\lambda}$ of the cooperon defined in
(\ref{eigenfunctions}), we obtain
\begin{eqnarray}\label{sigmaPerpTr}
  \frac{\Delta \sigmaBparCperp}{\sigmaZeroCperp}=-\frac{1}{\pi N_0 V}
  \sum_\lambda \frac{\bra{\Phi_\lambda} \cos a \hat{Q}_z
    \ket{\Phi_\lambda}}{E_\lambda}.
\end{eqnarray}
For the gauge $\vec{A}=(0,-B z,0)$ we can interpret the cosine as a
translation operator by one lattice site. In position space (note that
we are considering a lattice in the $z$-direction) the overlap is
therefore given by
\begin{eqnarray}
  \label{overlapCos}
  \bra{\Phi_\lambda} \cos a \hat{Q}_z  \ket{\Phi_\lambda}&=&
  \sum_n \int dx \,dy \nonumber \\
  && \hspace{-3em}
  \mal {\,\rm Re} \left[ \Phi^*_\lambda(x,y,a n+a)
    \Phi_\lambda(x,y,a n)\right].
\end{eqnarray}
where $n$ sums over the planes with distance $a$.

In the previous section we have shown that for low quantum numbers,
the eigenfunctions $\ket{\Phi_\lambda}$ can be approximated by
HO-eigenfunctions: $\Phi_{q_x,q_y,n}(x,y,z)\approx e^{i q_x x}
\Psi_n^{\text{HO}}(z- L_{m}^{2}q_{y}/2)$. In this regime $a$ is small
compared to the scale $\Delta_n$ on which $\Psi_n^{\text{HO}}(z)$
varies and $\bra{\Phi_\lambda} \cos a \hat{Q}_z
\ket{\Phi_\lambda}\approx 1+O((a/\Delta_n)^2)$. The scale $\Delta_n$
is of the order of half the distance of the nodes in
$\Psi_n^{\text{HO}}(z)$, i.e.
$\Delta_n\approx\sqrt{\Dperp/(n\Omegapar)}$.

For $a \gtrsim \Delta_n$ the overlap $\bra{\Phi_\lambda} \cos a
\hat{Q}_z \ket{\Phi_\lambda}$ vanishes on average. The condition for
the crossover $a \approx \Delta_{n_c}$, i.e.  $n_c \approx
2\Dbar/\wpar$ coincides with our previous estimate of the crossover to
the localized regime which is described by (\ref{locLoes}). As the
eigenfunctions of (\ref{locLoes}) are strictly localized on a single
plane, it is consistent to approximate $\bra{\Phi_\lambda} \cos a
\hat{Q}_z \ket{\Phi_\lambda}=0$ for $n > n_c$.

Within these approximations the only contribution to
(\ref{sigmaPerpTr}) stems from the HO regime. This contribution has
already been given in (\ref{K1}) and therefore
\begin{eqnarray}
  \frac{\Delta \sigmaBparCperp}{\sigmaZeroCperp}&\approx& -\lambda K_1 \\
  &\approx&
  -\lambda \left\{
    \begin{array}{l@{\quad}l}
      \alphaCperp - \betaBpar \wpar^{2}, & \wpar \ll x,\Dbar \\
      \alphaCperp - c \sqrt{{\wpar}/{\Dbar}},
      & x \ll \wpar \ll \Dbar \\
    \sqrt{\frac{\wpar}{2\Dbar}}
      g_{c}\left(\frac{\wpar}{4\Dbar}\right),
      & \Dbar,x \ll \wpar\\
      \propto 1/x, & x \gg 1
    \end{array}    
  \right. 
\end{eqnarray}
with
\begin{eqnarray}
  &K_1(\wpar \to 0)\approx \kappa_{2}(x+2\Dbar)-\kappa_{2}(x)\nonumber&\\
  &\displaystyle\kappa_{2}(z)=\frac{1}{\pi\sqrt{\Dbar}}\left[
    2\sqrt{z}\arctan\frac{1}{\sqrt{z}}+\log{(1+z)}\right].&
  \label{eq:kappaTwo}
\end{eqnarray}
Note that $K_1(\wpar \to 0)$ does not coincide exactly with
$\alphaCperp$ given in (\ref{eq:alphaCperp}), this is due to the
approximation $\bra{\Phi_\lambda} \cos a \hat{Q}_z
\ket{\Phi_\lambda}\approx 1$ which slightly influences the
contributions from higher energies (this explains the
small offset in Fig.~\ref{fig:2} for $\wparperp \to 0$). The
qualitative behavior, especially for low temperatures is however not
affected.

As $K_2$ is independent and $K_{3}$ only a smooth function of the
magnetic field, the magnetoresistance in a field parallel to the
planes is more or less independent of the direction of the applied
voltage. The main difference of $\Delta
\sigmaBparCperp/\sigmaZeroCperp$ to $\Delta
\sigmaBparCpar/\sigmaZeroCpar$ is the magnetic field independent, but
temperature dependent contribution $K_2$.

Our analysis of (\ref{overlapCos}) clearly indicates that the coherent
transport between {\em different} planes is essential for the WL
correction to the out-of-plane conductivity. This has e.g. the
consequence that for a strong magnetic field parallel to the plane
($\wpar\gg\Dbar$) the WL correction to the out-of-plane conductivity is
totally suppressed, while the conductivity parallel to the planes is
still affected by interference effects of electrons moving in a plane.

\section{Quasi One-Dimensional Systems}
Now, we will briefly discuss the quasi $1D$-case, concentrating on a
magnetic field perpendicular to the chains. We choose the symmetry
axis in the $z$-direction so that the analog of (\ref{cooperon2}) reads
\begin{equation}
  \left[ \Dpar \operator{Q}_{\|}^2 
    +\frac{4\Dperp}{a^2} \sum_{i=x,y}
      \sin^{2} \frac{a\operator{Q}_i}{2}
    +\frac{1}{\tauphi}\right] C(\vec{r})
  =\frac{1}{\tauel} \delta(\vec{r}).
  \label{cooperon4}
\end{equation}
A straightforward calculation similar to the calculation for the
quasi two-dimensional case with a magnetic field along the
$x$-axis and vector potential $A_{\perp}=(0,0,By)$ yields (using the
appropriate density of states $N_{0}^{1d} \approx (\pi \vf
a^{2})^{-1}$) the following quantum correction to the conductivity for
a current parallel to the chains
\begin{equation}
  \frac{\Delta \sigma_{\vec{B}\perp\unitvec{z}}^{1d}}{\sigma_0} 
  \approx  
  -\frac{K_{1}^{1d}+ K_{2}^{1d}- K_{3}^{1d}}{4\pi\sqrt{\Dbar}} 
\end{equation}
where $K_{1,2,3}^{1d}$ are given by 
\begin{eqnarray}
  K_{1}^{1d}&=&  \sum_{n=0}^{\infty}
  \frac{\wpar \,
    g_c\left[\left(n+\frac{1}{2}\right)\frac{\wpar}{2\Dbar}\right]}{%
    \sqrt{\epsilon_{n}+x}\sqrt{\epsilon_{n}+x+4\Dbar}}
  \label{eq:K1:1d}\\
  K_{2}^{1d}  &=& 4 \sqrt{\Dbar}
  \int_{0}^{\infty} \frac{g_{c}(q^{2})\,dq}
  {\sqrt{2\Dbar +x+q^{2}}\sqrt{6\Dbar +x+q^{2}}}
  \label{eq:K2:1d}\\
  &\approx& 
  \left\{
    \begin{array}{l@{\quad}l}
      \tilde{c}-4\sqrt{\Dbar}, &  x \ll \Dbar\\
      2\pi\sqrt{\Dbar/x} & \Dbar \ll x \ll 1\\
      4\sqrt{\Dbar}/x & x \gg 1
    \end{array}  
    \right.\\
  K_{3}^{1d}&=&  \sum_{n=0}^{\infty}
  \frac{\wpar \,
    g_c\left[\left(n+\frac{1}{2}\right)\frac{\wpar}{2\Dbar}\right]}{%
    \sqrt{x+2\Dbar}\sqrt{x+6\Dbar}}
  \label{eq:K3:1d}
\end{eqnarray}
with $\epsilon_{n}=(n+\frac{1}{2})\wpar$ and a numerical constant
$\tilde{c}\approx 3.31$. Note, that for the numerical evaluation of
(\ref{eq:K1:1d}), the lowest energy value $\epsilon_{0}$ should be
treated special according to (\ref{eq:epsilon:Zero}).

The limiting behavior of $\Delta\sigma$ in the various regimes is
given by
\begin{equation}
 \frac{\Delta \sigma_{\vec{B}\perp\unitvec{z}}^{\|, 1d}}{\sigma_0} 
 \approx \frac{-1}{4\pi\sqrt{\Dbar}}
 \left\{
   \begin{array}{l@{\quad}l}
     \alphaCpar_{1d}-\betaBpar^{1d}\wpar^{2}, & \wpar\ll x,\Dbar\\
     \alphaCpar_{1d}-c\sqrt{{\wpar}/{\Dbar}}, 
     & x \ll \wpar \ll \Dbar \\
     \displaystyle K_{2}^{1d} , 
     & \Dbar \ll \wpar \\
     \propto 1/x, & x\gg 1\\
   \end{array}
 \right. 
\end{equation}
with $c \approx 0.30$ and
\begin{eqnarray}
  \alphaCpar_{1d}&\approx& \frac{2\sqrt{\Dbar}}{\pi}
  \int_{0}^{\sqrt{2}\pi} dq_{\perp} 
  \int_{0}^{\infty} dq_{z} 
  \frac{q_{\perp} \, g_{c}(q_{\|}^{2})}{q_{\|}^{2}+\Dbar q_{\perp}^{2}+x}\\
  &=&\kappa_{2}(x+2\pi^{2}\Dbar)-\kappa_{2}(x) \nonumber\\
  \betaBpar^{1d}&=&-(2\Dbar+z)\lambda_{1}(z)\Big|_{z=x}^{x+2\Dbar}
\end{eqnarray}
where $\kappa_{2}$ can be found in (\ref{eq:kappaTwo}) and $\lambda_{1}$ in
(\ref{eq:lambdaOne}).

$K_{1}^{1d}$ can -- up to a changed cutoff and a different scaling of
the magnetic field -- be identified with (\ref{summ1}).
$\alphaCpar_{1d}$ was directly calculated in a vanishing magnetic
field using the approximation $2\sin^{2}(x/2) \approx x^2/2$ and the
cutoff $q_{x}^{2}+q_{y}^{2}=q_{\perp}^{2} < 2\pi^2/a^{2}$.

Again we find a term which is independent of the magnetic field, which
dominates in strong fields. This contribution remains, since the motion
of the electrons along the chains is not affected by the magnetic
field.  As expected for a quasi one-dimensional system, a dimensional
crossover from three-dimensional behavior at low temperatures, i.e.
small $x$, and small magnetic field to a one-dimensional one at higher
temperatures and fields can be observed.

In zero field and for low temperatures the contribution of WL to the
conductivity is finite and proportional to a $\sqrt{x}$, which is
typical for three dimensions. If the phase-coherence length is shorter
than the lattice distance ($\Dbar<x$), the correction to the
conductivity is proportional to $\alphaCpar_{1d}\approx
\pi^{2}\sqrt{\Dbar/x} +O\left[x,(\Dbar/x)^{3/2}\right]$ -- signalizing
the one-dimensional regime. A similar picture arises at low
temperatures as a function of the magnetic field: for a small magnetic
field and $x \ll \Dbar$ the system shows typical $3D$ behavior with a
$B^2$ contribution to the magnetoconductivity crossing over to a
$\sqrt{B}$ dependence as observed now for all systems discussed in the
paper.  But at strong enough fields, $\wpar \approx \Dbar$, a
crossover to the one-dimensional case can be seen, where WL is not
influenced by a magnetic field and the finite contribution
$K_{2}^{1d}$ remains.

For a current perpendicular to the chains, the situation is similar
to the quasi $2D$ case: $K_{1}^{1d}$ is unchanged whereas the
localized regime $K_{2}^{1d}-K_{3}^{1d}$ does not contribute.
\begin{eqnarray}
  \frac{\Delta \sigma_{\vec{B}\perp\unitvec{z}}^{\perp, 1d}}{\sigma_0} 
  &\approx&  -\frac{K_{1}^{1d}}{4\pi\sqrt{\Dbar}} \nonumber\\
 &\approx& \frac{-1}{4\pi\sqrt{\Dbar}}
  \left\{
    \begin{array}{l@{\quad}l}
      \alphaCperp_{1d}-\betaBpar^{1d}\wpar^{2}, & \wpar\ll x,\Dbar\\
      \alphaCperp_{1d}-c\sqrt{{\wpar}/{\Dbar}}, 
      & x \ll \wpar \ll \Dbar \\
      2 g_{c}\left({\wpar}/{4\Dbar}\right), 
      & x \ll \Dbar \ll \wpar \\
      \propto 1/x, & x\gg 1\\
    \end{array}
  \!\!\!\right.
\end{eqnarray}
with
\begin{equation}
  \alphaCperp_{1d}=\kappa_{1}(x+2\Dbar)-\kappa_{1}(x).
\end{equation}
$\kappa_{1}$ is defined in (\ref{eq:kappaOne}).  Here for
$x>\Dbar$ or $\wpar>\Dbar$ weak localization is strongly suppressed.

\section{Universality and Experiments}
Recently Zambetaki, Li {\it et al.} \cite{Zambetaki:96} have
numerically analyzed a quasi two-dimensional system and found that
one-parameter scaling can still be applied and that such an
anisotropic system is in the universality class of an anisotropic
three-dimensional system.  In this paper we have emphasized that the
distance of the planes $a$ is an important extra length scale and that
the corresponding dimensionless quantity $\Dbar$ governs the physics
of the dimensional crossover. Nevertheless for a weak magnetic field
$x, \wparperp \ll \Dbar$ we recover universal behavior as has to be
expected \cite{Lee:85}.

\begin{figure}
  \begin{center}
    \epsfig{width=\linewidth,file=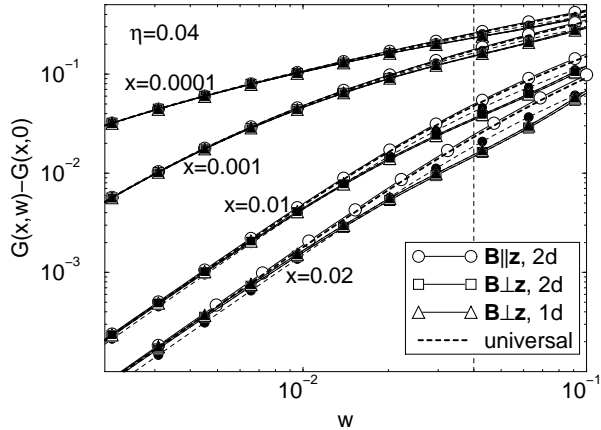}
    \caption{
      Magnetoconductivity scaled according to (\ref{universalG}) 
      for various temperatures. We investigate both quasi $1D$- and
      quasi $2D$-systems.  The open and filled symbols denote the
      conductivity parallel and perpendicular to the planes,
      respectively.  For low magnetic field and low temperatures the
      effects of WL are universal and do not depend on $\Dbar$ or the
      structure of the material, corrections are of the order of
      $w/\Dbar$ or $x/\Dbar$.  The thick dashed line is the universal
      result for a three-dimensional system. All curves of the plot
      are given for $\Dbar=0.04$. The universal behavior for $x,w \ll
      \Dbar$ is independent of $\Dbar$.}
    \label{fig:4}
  \end{center}
\end{figure}

Actually, we think that exploiting the universality at low
temperatures and magnetic field might serve as a valuable tool to
analyze experimental results. E.g. leading corrections to the
universal behavior show up in powers of $\wparperp\,/\Dbar$ or
$x/\Dbar$ which allows the determination of $\Dbar$ but also of
$x=\tauel/\tauphi$.

The absolute size of the WL-correction is for a three-dimensional
system always non-universal, i.e. it depends on the behavior of the
system on the short length scale $\lel$. A more well-suited quantity
is e.g.  $(\sigma(T,\vec{B})-\sigma(0,0))/\sigma(0,0)$ which for a
generic anisotropic three dimensional system with diffusion constants
$D_x$, $D_y$, $D_z$ has for low temperatures and low magnetic field the
generic form:
\begin{eqnarray}
  \label{universalG}
  \frac{\sigma(T,\vec{B})-\sigma(0,0)}{\sigma(0,0)}
  &=&
  \frac{\tauel}{\pi N_0 \sqrt{D_x D_y D_z \tauel^3}} \,G(x,w)
  \label{universalG2}\\
  &=&
  \frac{1}{4 \pi^2 N_0 \sqrt{D_x D_y D_z}}\, \sqrt{\frac{x}{\tauel}} 
  g \left(\frac{w}{x}\right)
\end{eqnarray}
\begin{equation}
  w^2=\left(\frac{2 e \tauel}{c}\right)^2
  \left(D_x D_y B_z^2+D_y D_z B_x^2+D_z D_x B_y^2\right)
\end{equation}
$g(z)$ is a universal function \cite{Kawabata:80a} with
$g(0)=1$ and $g(z) \to (\zeta(1/2)/2)(1-\sqrt{2}) \sqrt{z}\approx 0.3024
\sqrt{z}$ for $z \to \infty$. Note that (\ref{universalG2}) is independent
of the cutoff scale $\tauel$.

It is easy to check that for $\wparperp , x \ll \Dbar$ {\em{all}} our
results for all directions of the current and the magnetic field, and
for both the quasi one- and two-dimensional case can be written in
this way. In Fig.~\ref{fig:4} we have scaled the magnetoconductivity
in the above described way -- all curves collapse on a single line for
$w,x \ll \Dbar$, signalizing the dominating three dimensional
behavior. With the help of (\ref{universalG2}) it is also possible to
scale the universal part to a {\em single} line independent of $x$ and
$\Dbar$.

In an experiment it would be very interesting to study systematically
the deviations from the $3D$-universality. In the universal $3D$
regime the WL corrections do not depend on the direction of the
current and the dependence on the direction of the magnetic field is
fully described by the $\vec{B}$-dependence of $w$.  Therefore
deviations are studied best looking at ratios of conductivities. E.g.
without a magnetic field for low temperatures one could investigate
the ratio
  \begin{equation}
    C(T)=\frac{\sigmaCpar(T)-\sigmaCpar(0)}{\sigmaCperp(T)-\sigmaCperp(0)}.
  \end{equation}
  With $C(0)=\lim_{T \to 0} C(T)$ the ratio $C(T)/C(0)\approx
  1+\sqrt{x/(4 \Dbar)}+O(\sqrt{x \Dbar}, x/\Dbar)$ calculated from
  (\ref{eq:alphaCpar}) and (\ref{eq:alphaCperp}) for a quasi $2D$
  system measures the leading deviations from universality.  Similar
  information, perhaps with an higher accuracy, one can get from the
  $B^2$-rise of the magnetoconductivity, i.e. from
  \begin{equation}
    D(T)=\left.  \frac{\frac{d^2\hfill}{dB^2} \sigmaBperpCpar(T)}
      {\frac{d^2\hfill}{dB^2} \sigmaBperpCperp(T)} \right|_{B=0}.
\end{equation}
Using the extrapolation of $D(T)$ towards $T=0$ with $D(0)=C(0)$, one
can investigate the behavior of $D(T)/D(0)\approx 1-x/(2
\Dbar)+O(\sqrt{\Dbar x^3},(\Dbar/x)^2)$. This correction can be seen
in the $\log$-$\log$ plot of the magnetoresistivity shown in
Fig.~\ref{fig:4}, where for $w\to 0$ the curves show a small offset
proportional to $x$.

Corrections to the universal $3D$ behavior proportional to powers of
$w/\Dbar$ are harder to investigate systematically. For $w<x$ it is
difficult to separate them from $x/\Dbar$ contributions and the regime
$w>x$ is very sensible to small variations of the temperature.  In a
quasi $2D$ system we propose to compare the in-plane and out-of-plane
conductivity for a magnetic field perpendicular to the planes using
  \begin{equation}
    E(T,B)=
    \frac{\sigmaBparCpar(T,B)-\sigmaBparCpar(T,0)}%
    {\sigmaBparCperp(T,B)-\sigmaBparCperp(T,0)}
\end{equation}
as theoretical uncertainties are minimal for this quantity. $E(T,B)$
rises quadratically in $B$ for low fields, but for low
temperatures it should be possible to extract a {\em linear}
contribution in $B$ in the regime $x<\wperp<\Dbar$.  For $T \to 0$
this rise is proportional to $\wperp/\Dbar$ allowing a measurement of
$\Dbar$.  We find $E(0,B)/E(0,0)\approx 1+0.025
(\wpar/\Dbar)(1+O(\sqrt{x/\Dbar})) +O(x/\Dbar)$.

The predictions of our paper are relevant for a number of
experimentally available systems.  In the quasi one-dimensional case,
one has to look for systems which stay metallic at low temperatures
and do not exhibit a Peierls' transition or a transition to a
superconducting state. This seems to be realized in certain pure
iodine oxidized phthalocyanine molecular crystals ($M(pc)I$ with
$M=H_2,Ni, Cu, \dots$) where indeed a dimensional crossover as a
function of temperature has been observed \cite{Lee:96}. However,
measurements of the magnetoresistance did not show deviations from the
$B^2$ behavior, which would be necessary to analyze the data with our
theory. Here investigations at lower temperatures and higher magnetic
fields would hopefully allow to test our picture.

Very promising seems to be the investigation of quasi two-dimensional
systems which do not show a phase-transition and stay metallic at low
temperatures.  Especially metal-insulator multilayers can be
fabricated in a well-controlled way. All parameters entering our
theory, i.e.  strength of disorder, anisotropy, bandstructure etc. can
be be varied over a large range. A number of experiments in such
multilayers \cite{Jin:86,Baxter:96} have shown signatures
of a dimensional crossover from two- to three-dimensional behavior. It
is important to study systems with quite strong disorder where WL is
enhanced and contributions from other effects like classical
magnetoresistance, Shubnikov-de Haas oscillations are suppressed or
can be separated.

Especially the comparison of the effect of a magnetic field parallel
and perpendicular to the planes would allow to identify e.g. the
magnetic field independent contribution $K_2$.

Finally we want to give a crude estimation for the required magnetic
fields necessary to induce a dimensional crossover in a quasi $2D$
system.  $\wpar \gg \Dbar$ in our dimensionless units translates to $B
\gg \sqrt{\Dperp/\Dpar} \phi_0/(4 \pi a^{2})$ with the flux-quantum
$\phi_0=h/2e=2.07 \mal 10^{-15} \, \mathrm{T}\, \mathrm{m}^2$.  E.g.
for $D_\perp/D_\|\sim 10^{-3}$ and $a\sim \, 10 \, \mathrm{\AA}$ this
yields $B \gg 5 \,\mathrm{T}$. The field perpendicular to the planes
can be a factor $\sqrt{D_\perp/D_\|}\approx 30$ smaller.  In
multilayer materials much larger values for $a$ can be achieved and
correspondingly much weaker magnetic fields should lead to the
dimensional crossover discussed in this paper. Note that $a$ enters
our estimate quadratically.

\section{Conclusions}
In this paper we have calculated the dimensional crossover in WL
induced by inelastic scattering or a magnetic field, for anisotropic
three-dimensional systems. At low temperatures and small magnetic
field we recover the results for an anisotropic three-dimensional
system as different planes or chains are connected by coherent
diffusion processes.  However, with increasing temperature, as the
phase coherence length gets shorter than the lattice distance, a
crossover to the two- or one-dimensional behavior of WL can be
observed. A similar effect can be seen for an increasing magnetic
field. If a typical diffusion path connecting different planes of
chains encloses a flux quantum, coherence is destroyed and a
dimensional crossover is induced.  This phenomenon depends crucially
on the direction of both the applied field and the current.  We
propose to use the magnetic field dependence as a tool to investigate
these quantum-interference effects and the dimensional crossover in
detail.  As compared to the temperature dependence, the magnetic field
has the advantage not to dependent on the uncertainties associated
with the phase relaxation mechanism. A main qualitative result of our
calculations is that for a magnetic field parallel to the planes of a
quasi two-dimensional system WL is not fully suppressed by a magnetic
field.  A finite contribution remains which has its origin in
diffusion processes in the plane which do not enclose magnetic flux.
It should be possible to measure this remaining contribution by
comparing the magnetoresistance for a magnetic field parallel and
perpendicular to the planes.

A generalization of this theory to crossovers e.g. from $D=1$ to $D=2$
is straightforward. We have not discussed the influence of WL on the
frequency-dependent conductivity \cite{Sumanasekera:94} in this paper.
The effect of a finite frequency $\omega$ in a microwave experiment
($\omega \tauel \ll 1$) can easily be included by replacing
$x=\tauel/\tauphi$ by $x-i \omega \tauel$ in our formulas. Such a
microwave experiment has the advantage that $\omega$ is a known
frequency while $1/\tauphi$ is only indirectly accessible.

\section{Acknowledgment}
We wish to thank Yoonseok Lee and W.~P.~Halperin for helpful
communications.  C. M. acknowledges financial support 
of the German-Israeli Foundation.

\end{document}